\documentclass[12pt]{iopart}

\expandafter\let\csname equation*\endcsname\relax

\expandafter\let\csname endequation*\endcsname\relax

\usepackage{amsmath}
\usepackage{amssymb}
\usepackage[mathscr]{eucal}

\usepackage{graphicx}
\usepackage{lipsum}

\usepackage{color}
\definecolor{dgreen}{rgb}{0,0.7,0}

\newcommand{\dd}{\mathrm{d}}
\newcommand{\ee}{\mathrm{e}}

\newcommand{\tc}{t_\mathrm{c}}
\newcommand{\cs}{c_\mathrm{s}}

\newcommand{\xipm}{\xi_\pm}
\newcommand{\Tp}{T_+}
\newcommand{\Tm}{T_-}

\newcommand{\Tb}{\overline{T}}
\newcommand{\Js}{J^\mathrm{stat}}

\newcommand{\cTp}{\mathcal{U}}

\newcommand{\cTpfixed}{\cTp^\text{fixed}}
\newcommand{\cTpinf}{\cTp^\infty}
\newcommand{\cK}{\mathcal{K}}

\newcommand{\cKinf}{\cK^\infty}
\newcommand{\cJ}{\mathcal{J}}

\newcommand{\phip}{\phi_+}
\newcommand{\phim}{\phi_-}
\newcommand{\phipm}{\phi_\pm}
\newcommand{\phiz}{\phi_0}
\newcommand{\etap}{\eta_+}
\newcommand{\etam}{\eta_-}
\newcommand{\etaz}{\eta_0}
\newcommand{\etapm}{\eta_\pm}
\newcommand{\Dz}{D_0}

\begin{document}

\title{Temperature profile and boundary conditions in an anomalous heat transport model}

\author{J Cividini$^1$, A Kundu$^2$, A Miron$^1$ and D Mukamel$^1$}
\address{$^1$ Department of Physics of Complex Systems, Weizmann Institute of Science, Rehovot 76100, Israel}
\address{$^2$ International center for theoretical sciences, TIFR, Bangalore - 560012, India}
\ead{julien.cividini@weizmann.ac.il}
\vspace{10pt}

\begin{abstract}
A framework for studying the effect of the coupling to the heat bath in models exhibiting anomalous heat conduction
is described. The framework is applied to the harmonic chain with momentum exchange model where the non-trivial temperature profile is calculated. In this approach one first uses the hydrodynamic (HD) equations to calculate
the equilibrium current-current correlation function in large but finite chains, explicitly taking into account the BCs
resulting from the coupling to the heat reservoirs.  Making use of a linear response relation, the anomalous conductivity exponent $\alpha$ and an integral equation for
the temperature profile are obtained. The temperature profile is found to be singular at the boundaries with an exponent which
varies continuously with the coupling to the heat reservoirs expressed by the BCs. In addition, the relation between
the harmonic chain and a system of noninteracting L{\'e}vy walkers is made explicit, where different BCs of the chain correspond to
different reflection coefficients of the L{\'e}vy particles.
\end{abstract}

\vspace{2pc}

\vspace{2pc}
If one imposes two temperatures $\Tp$ and $\Tm$ at the left and right ends of a large 1D system of length $N$, a naive application of Fourier's law predicts a heat current $J = \kappa \frac{\Tp-\Tm}{N}$, where $\kappa$ is an $N$-independent quantity called the conductivity. Anomalous heat conduction, typically characterized by $\kappa = a N^\alpha$ with $0<\alpha\leq1$, has however been shown to be a generic feature of one-dimensional momentum-conserving systems~\cite{lepri_l_p2003, dhar2008, lepri2016}. Violations of Fourier's law have been reported for a large variety of systems over the last decades both in theoretical models~\cite{lepri_l_p1997, grassberger_n_y2002, mai_d_n2007, das_d_n2014, hurtado_g2016, xiong2016, dhar2001a, casati_p2003, mendl_s2014, prosen_c2000, narayan_r2002, basile_b_o2006, jara_k_o2015, komorowski_o2016, bernardin_g_j2016} and in nanotube experiments~\cite{chang2008}.
This phenomenon is usually accompanied by other puzzling features such as divergence of the time integral of the equilibrium current-current correlations, superdiffusive propagation of local energy perturbations~\cite{liu2014} and boundary singularities in the nonequilibrium stationary temperature profile \cite{lepri2016}.

For an infinite system, the heat conductivity is generally obtained from the equilibrium current-current correlation function using the Green-Kubo formula~\cite{kubo_t_h1991, lepri2016},
\begin{equation}
\kappa = \lim_{\tau\to\infty} \lim_{N\to\infty} \frac{1}{ \Tb^{2}N}\int_{0}^\tau \hspace{-2mm}\dd t \iint_{[0;N]^2} \hspace{-6mm} \dd x \dd y \langle J(x,t) J(y,0) \rangle_{N,\mathrm{eq}}. \label{eq:greenkubo}
\end{equation}
Here $J(x,t)$ is the instantaneous energy current, $\Tb = \frac{\Tp+\Tm}{2}$ is the average temperature, $\langle \ldots \rangle_{N,\mathrm{eq}}$ denotes the equilibrium average and we work in units where $k_B=1$  throughout the paper. 

In case of anomalous conduction \eqref{eq:greenkubo} diverges and is not applicable in a strict sense. Arguing that correlations are strongly suppressed at time scales when the sound peaks hit the boundaries, a cutoff time $\tc \propto N$ is introduced~\cite{narayan_r2002}. This cutoff causes the effective conductivity to depend on $N$ and to diverge for $N\to\infty$. 

Anomalous conductivities have been observed to be accompanied by temperature profiles with divergent derivatives at the boundaries~\cite{lepri_l_p1997, dhar2001a, aoki_k2001, casati_p2003, lepri_l_p2003, mai_d_n2007, delfini2008, dhar2008, lepri_m_p2009, lepri_m_p2010, delfini2010, gerschenfeld_d_l2010, lepri_p2011, das_d_n2014, lepri2016, hurtado_g2016}, namely $\frac{\dd T}{\dd x} \sim_{x \to 0^+} x^{-\nu}$ with $0<\nu<1$ and symmetrically for $x\to N^-$. An exact calculation of the temperature profile has been carried out for the harmonic chain with momentum exchange (HCME) in fixed boundary conditions (BCs)~\cite{lepri_m_p2009}, but a more general theoretical framework is lacking. 
In a related phenomenological approach to heat conduction, quanta of heat are assumed  to undergo L{\'e}vy flights~\cite{blumen_z_k1989, denisov_k_u2003, cipriani_d_p2005, zaburdaev_d_h2011, lepri_p2011,dhar_s_d2013}. The L{\'e}vy particle density and current hence become analogous to the temperature and the energy current, respectively. Excellent agreement has been observed in some cases~\cite{lepri_p2011}, but there is no known generic way to translate a heat conducting system into a L{\'e}vy equivalent.

In this work we present a general framework which takes into account the effect of BCs for evaluating temperature profiles and in particular the exponent $\nu$. We then apply this method to the HCME.
Let us start by introducing some general considerations. 
For small temperature differences $\Delta T = \Tp-\Tm \ll \Tb$, the stationary distribution is close to local equilibrium. 
Linearizing around local equilibrium as was done in~\cite{kundu_d_n2009}, we obtain~\cite{longpaper},  
\begin{equation}
\label{eq:linres}
\Js = -\int_{y=0}^N K_N(x,y) \frac{\dd T}{\dd y}(y) \dd y,
\end{equation}
where $\Js$ is the stationary heat current, which does not depend on space. The kernel $K_N(x,y)$ is expected to be determined by the equilibrium properties of the system through
\begin{equation}
\label{eq:defK}
K_N(x,y) = \frac{1}{\Tb^2} \int_{t=0}^\infty \langle J(x,t) J(y,0)\rangle_{N,\mathrm{eq}} \dd t.
\end{equation}
At this point, replacing the current-current correlation by its infinite system counterpart generally (although not always~\cite{dhar2001b}) provides the correct scaling exponent $\alpha$ after properly introducing the upper cutoff time $t_c$ in \eqref{eq:greenkubo}. 
There is however no guarantee that this approximation can be made in general. In several studies the temperature profile~\cite{delfini2010}, the current prefactor $a$~\cite{rieder_l_l1967, aoki_k2001, delfini2010} or even the exponent $\alpha$~\cite{matsuda_i1970, keller_p_w1978, dhar2001b} were found to depend on the BCs.

In the context of anomalous transport, one can 
assume a large-$N$ scaling form of the temperature profile,
\begin{equation}
\label{eq:scalingTprofJ}
-\frac{N}{\Delta T} \frac{\dd T}{\dd y}(y) \to \cTp\left(\frac{y}{N}\right),~\frac{N^{1-\alpha}}{\Delta T} \Js \to \cJ
\end{equation}
where $\cTp$ and $\cJ$ do not depend on $N$ and $\Delta T$.
Using the scaling forms~\eqref{eq:scalingTprofJ} in the large $N$ limit, \eqref{eq:linres} becomes
\begin{equation}
\label{eq:linressc}
\int_{v=0}^1 \cK(u,v) \cTp(v) \dd v = \cJ,
\end{equation}
with the asymptotic kernel
\begin{equation}
\label{eq:scalingK}
\cK(u,v) = \lim_{N\to \infty} N^{1-\alpha}K_N(Nu,Nv),
\end{equation}
where the fact that the limit~\eqref{eq:scalingK} exists fixes the value of the exponent $\alpha$. Equation~\eqref{eq:linressc} determines both the temperature profile $\cTp(u)$ and the current $\cJ$ once the boundary values of the temperature are specified. Here we simply assume that there is no discontinuity at the boundaries, which imposes the normalization $\int_{u=0}^1 \cTp(u) \dd u = 1$. One can therefore solve~\eqref{eq:linressc} for $\cTp(u)$ up to a multiplicative constant, impose the normalization and then compute $\cJ$ using~\eqref{eq:linressc} with known $\cTp(u)$.

Assuming that $\cK(u,v)$ can be replaced by its infinite system expression $\cKinf(u,v) = \cKinf(|u-v|) = |u-v|^{\alpha-1}$, as imposed by the scaling~\eqref{eq:scalingK} and translational invariance, yields further simplification. Inverting~\eqref{eq:linressc} using the Sonin inversion formula~\cite{buldyrev2001, dhar_s_d2013} yields 
\begin{equation}
\label{eq:Tpsonin}
\cTpinf(u) = \frac{\Gamma[2-\alpha]}{\Gamma\left[ 1-\frac{\alpha}{2}\right]^2} [u(1-u)]^{-\frac{\alpha}{2}},
\end{equation}
from which the temperature profile is easily obtained by integration.
In the L\'evy flight picture, for absorbing boundaries the density profile precisely obeys Eq.\,\eqref{eq:linressc} with the infinite-system kernel~\cite{drysdale_r1998, buldyrev2001, dhar_s_d2013} and is therefore given by~\eqref{eq:Tpsonin}. 

To analyze the effect of the BCs, we consider the HCME model~\cite{delfini2008, lepri_m_p2009, lepri_m_p2010, delfini2010, jara_k_o2015, komorowski_o2016}, where the particles $i = 1,\ldots,N$ have positions $q_i(t)$ and momenta $p_i(t)$, and the left and right ends are coupled to Langevin reservoirs. 
The equations of motion for $i=1,\ldots,N$ are
\begin{eqnarray}
\label{eq:eombulk}
\frac{\dd q_i}{\dd t} &=& p_i, \nonumber \\
\frac{\dd p_i}{\dd t} &=& (1-\delta_{i,1}-\delta_{i,N}) \omega^2 (q_{i+1}-2 q_i +q_{i-1}) \\
&& + \delta_{i,1} \left[\omega^2 (q_2- \zeta q_1) + \xi_1-\lambda p_1\right]  \nonumber \\ && + \delta_{i,N} \left[\omega^2 (q_{N-1}-\zeta q_N) + \xi_N-\lambda p_N \right], \nonumber
\end{eqnarray}
where $\zeta=1$ for \emph{free} BCs and $\zeta=2$ for \emph{fixed} BCs.
Here $\omega$ is the oscillator frequency and 
$\xi_1$ and $\xi_N$ are two mean-zero Gaussian noises with variances $2 \lambda \Tp$ and $2 \lambda \Tm$ originating from the heat baths at temperatures $\Tp$ and $\Tm$ attached to the left and right ends respectively. 
In addition to the noisy dynamics~\eqref{eq:eombulk}, momenta of neighboring sites are exchanged with a constant rate $\gamma$ to allow the model to relax to a Gibbs state under equilibrium conditions. Here we are mainly interested in the temperature profile $T_i = \langle p_i^2 \rangle$ for $i=1,\ldots,N$.

The model has been analyzed analytically for fixed BCs and studied numerically for free BCs. It has been shown that the anomalous exponent $\alpha=\frac{1}{2}$ in both cases~\cite{delfini2008, lepri_m_p2009, delfini2010, spohn2014, spohn_s2015, mendl_s2015}.
For fixed BCs the asymptotic temperature profile is known exactly~\cite{lepri_m_p2009},
\begin{equation}
\label{eq:dTfixed}
\cTpfixed(u) = \frac{\pi \sqrt{2}}{(\sqrt{8}-1) \zeta(3/2)} \sum_{p=0}^\infty \frac{\sin (\pi (2 p+1)u)}{(2p+1)^{1/2}}
\end{equation}
and is independent of $\lambda$, whereas it has been shown numerically to depend on $\lambda$ for free BCs~\cite{delfini2010}. In particular,  the solution~\eqref{eq:dTfixed} diverges near both boundaries $u=0$ and $u=1$ with an exponent $\nu=\frac{1}{2}$. 
However, replacing $\cK$ by $\cK^\infty$ gives the solution~\eqref{eq:Tpsonin} and $\nu=\frac{1}{4}$, in contradiction to \eqref{eq:dTfixed}. One therefore needs to compute the kernel $\mathcal{K}$ in presence of the 
boundaries.

In order to do that we first translate the microscopic equations~\eqref{eq:eombulk} to hydrodynamic (HD) equations. In the bulk, a general HD approach has recently been developed~\cite{vanbeijeren2012, spohn2014, spohn2015, spohn_s2015, mendl_s2015, popkov2016}, which provides a general framework accounting for the anomalous behavior. On HD time scales a random fluctuation decomposes into two balistically moving sound peaks, whose evolution is governed either by Kardar-Parisi-Zhang or Edwards-Wilkinson equations, and one symmetric heat peak generically shaped like a L{\'e}vy distribution~\cite{spohn2014,lepri2016}. We thus define the stretch and energy variables $s_i = q_{i+1} - q_i$ and $e_i = \frac{p_i^2}{2} + \omega^2 \frac{s_{i}^2 + s_{i-1}^2}{4}$ and coarse-grain the equations of motion~\eqref{eq:eombulk} in space and time. Expressing the equations in terms of the sound modes $\phipm = \omega s \mp p$ and heat mode $\phiz=e$, and adding diffusion and noise terms, 
as prescribed by the fluctuating HD framework~\cite{vanbeijeren2012, bernardin_o2014, spohn2014, spohn_s2015, mendl_s2015, popkov2016, lepri2016}, one obtains~\cite{spohn2015, dhar_s_r2015, lepri2016}
\begin{eqnarray}
\label{eq:hydro}
\partial_t \phipm &=& - \partial_x [\pm \cs \phipm - D \partial_x \phipm - \sqrt{2 D} \etapm], \\
\partial_t \phiz &=& - \partial_x [ G(\phip^2-\phim^2) - \Dz \partial_x \phiz - \sqrt{2 \Dz} \etaz], \nonumber
\end{eqnarray}
where $\cs = \omega$ is the speed of sound, $\etap$, $\etaz$
and $\etam$ are uncorrelated Gaussian white noises, 
$G = \frac{\omega}{4}$ results from the nonlinearity 
and $D$ and $\Dz$ are phenomenological diffusion coefficients. The momentum exchange mechanism ensures that the HCME thermalizes but may only appear in~\eqref{eq:hydro} through the diffusion constants $D$ and $\Dz$.

The instantaneous energy current can be read from~\eqref{eq:hydro},
\begin{equation}
\label{eq:defJ}
J(x,t) = G (\phip(x,t)^2-\phim(x,t)^2),
\end{equation}
neglecting the subdominant diffusion term~\cite{spohn2015}.
The fluctuating HD equations~\eqref{eq:hydro},\eqref{eq:defJ} have been shown to predict the right scaling for the conductivity~\cite{lepri2016} and the cumulants of the equilibrium current of the HCME on a ring~\cite{dhar_s_r2015}. Since neither the current~\eqref{eq:defJ} nor the equations for $\phip$ and $\phim$ depend on $\phiz$, the kernel~\eqref{eq:defK} can be obtained by solving only the equations for $\phip$ and $\phim$ with suitable BCs.

Contrary to the bulk equations, there is no general recipe for deriving the BCs for the HD equations.
Here the strategy is to introduce extra stretch and momentum variables in order to extend the structure of the bulk equations to the boundary equations $i=1,N$ at the cost of introducing additional conditions, that become the HD BCs after coarse-graining. As in the bulk, we expect the momentum exchange to lead to high order spatial derivatives, making it negligible after coarse-graining, see~\eqref{eq:hydro} and~\cite{bernardin_o2014}. The noises $\xipm$ do not contribute either since their time averages vanish.

\begin{figure}[t]
	\begin{center}
		\includegraphics[width=0.85\textwidth]{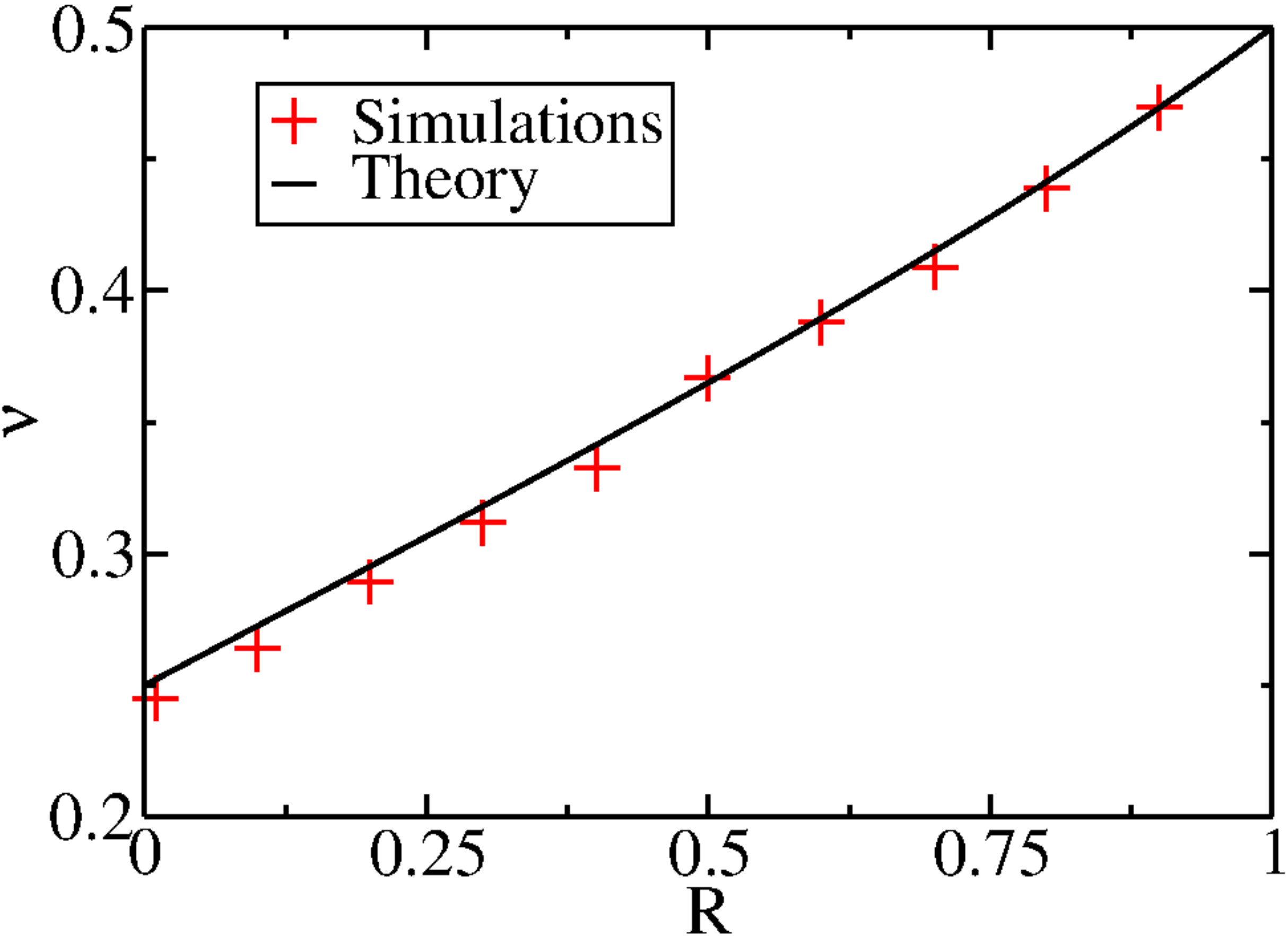} 
	\end{center}
	\caption{\small Exponent $\nu$ as a function of $R$. The prediction from~\eqref{eq:nuimpl} is plotted against the values obtained from the numerical solution of the integral equation~\eqref{eq:linressc}-\eqref{eq:solK}. }
	\label{fig:Rnu}
\end{figure}

To explain the procedure let us consider the free BCs case as an example. 
We introduce a dynamical variable $s_0$ such that $p_1$ follows the bulk evolution equation.
Neglecting the noise in \eqref{eq:eombulk}, we obtain the condition 
\begin{equation}
\omega^2 s_0 = \lambda p_1.
\end{equation}
A second BC is needed, since the HD equations~\eqref{eq:hydro} are of second order in space, and is obtained by introducing $p_0$ such that $s_0$ follows a regular equation of motion. Taking a time derivative of the first condition yields 
\begin{equation}
p_1 - p_0 = \lambda (s_1 - s_0).
\end{equation}
After coarse-graining space and expressing $s$ and $p$ in terms of the sound modes $\phipm$, 
we obtain for {\it free} BCs the HD BCs
\begin{eqnarray}
\label{eq:bcsfree}
(\partial_x \phip + r \partial_x \phim) |_{x=0} = ( \phip - r \phim) |_{x=0} &=& 0, \nonumber \\
(\partial_x \phim + r \partial_x \phip) |_{x=N} = ( \phim - r \phip) |_{x=N} &=& 0,
\end{eqnarray} 
where the condition at $x=N$ has been obtained by applying the same treatment to the equation for $p_N$, and
\begin{equation}
\label{eq:r}
r=\frac{\lambda-\omega}{\lambda+\omega}.
\end{equation}
Physically, the BCs~\eqref{eq:bcsfree} mean that when a $\phip$ (resp. $\phim$) Gaussian peak hits the right (resp. left) boundary, it turns into a $\phim$ (resp. $\phip$) Gaussian peak whose integral is multiplied by a factor $r$. Note that negative $r$ corresponds to a change of phase of the reflected peak relative to the incoming one. These phenomena have been observed in numerical simulations and the validity of \eqref{eq:bcsfree} has been confirmed (not shown here, see~\cite{longpaper}). The \textit{resonant} (impedance matching) case~$\lambda=\omega$ is of particular interest since there is no reflection at all, $r=0$.

The fixed BCs can be considered as a special case of the free BCs. Indeed, for $\lambda \to \infty$ the positions $q_1$ and $q_N$ stay very close to $0$ and therefore mimic fixed BCs. For fixed BCs we therefore have~\eqref{eq:bcsfree} with $r=1$. 

The bulk equations~\eqref{eq:hydro} and BCs~\eqref{eq:bcsfree} for $\phip$ and $\phim$ are linear and can be solved for any initial condition 
and reflection coefficient $r$. The solutions are written in terms of the four Green functions $f_{\sigma, \tau}(x,y,t)$ for $\sigma,\tau=\pm$, as 
\begin{eqnarray}
\label{eq:phipm}
&&\phi_\sigma(x,t) =  \sum_{\tau = \pm} \bigg[ \int_{y=0}^N \dd y f_{\sigma,\tau}(x,y,t) \phi_\tau(y,0)  \\ &&~+ \sqrt{2 D} \int_{y=0}^N  \dd y \int_{t'=0}^t \dd t' f_{\sigma,\tau}(x,y,t-t') \partial_{y} \eta_\tau(y,t') \bigg], \nonumber \\
&& \text{where,} \nonumber \\
&&	f_{\sigma,\tau}(x,y,t) = \sum_{n=-\infty}^{\infty} r^{2 n + \frac{\sigma-\tau}{2}} \frac{\ee^{-\frac{(x- \sigma \tau y + 2 \sigma n N -\sigma \cs t)^2}{4 D t}}}{\sqrt{4 \pi D t}}, \label{eq:greenf}
\end{eqnarray}
with $r=1$ for fixed BCs and $r=\frac{\lambda-\omega}{\lambda+\omega}$ for free BCs.

The expressions \eqref{eq:phipm}  for the fields can now be inserted in \eqref{eq:defJ} to compute the kernel~\eqref{eq:defK}. Averaging over the Gaussian initial conditions and integrating over time gives $K(x,y)$ (for details see~\ref{section:appkernel}). After large $N$ rescaling of the profile and the kernel~\eqref{eq:scalingTprofJ}-\eqref{eq:scalingK} we recover $\alpha = \frac{1}{2}$.
Up to corrections of order $N^{-1}$, the rescaled kernel reads
	\begin{eqnarray}
	\label{eq:solK}
	\cK(u,v) &=& \frac{G^2 S^2}{ \sqrt{2 \pi D \cs} \Tb^2} \bigg[ \frac{1}{\sqrt{|u-v|}} + \sum_{n=1}^\infty \bigg( \frac{R^{2 n}}{\sqrt{2 n+ u-v}} \nonumber \\ &&+ \frac{R^{2 n}}{\sqrt{2 n- u+v}} - \frac{R^{2 n-1}}{\sqrt{2 n-2 +u +v}} -  \frac{R^{2 n-1}}{\sqrt{2 n-u-v}} \bigg) \bigg],
	\end{eqnarray}
where $S = \langle \phip(x,0)^2\rangle_\mathrm{eq} = \langle \phim(x,0)^2\rangle_\mathrm{eq}$ denotes the amplitude of the equilibrium fluctuations, and $R=r^2$.
Solving~\eqref{eq:linressc} along with the kernel~\eqref{eq:solK} one obtains the temperature profile. Although our analysis is based on the assumption $|\Delta T| << \Tb$, for HCME this temperature profile will be valid for arbitrary $\Delta T$, since the quadratic correlations satisfy a closed set of linear equations with a source term proportional to $\Delta T$~\cite{lepri_m_p2009}.

Exact analytical expressions of the temperature profile can be obtained in two limiting cases. For the free resonant case $R=0$ the kernel is the same as that of an infinite system and depends only on $u-v$. The profile is therefore given by~\eqref{eq:Tpsonin} with $\alpha=\frac{1}{2}$, yielding an exponent $\nu=\frac{1}{4}$ for the boundary singularities. For the fixed case $R=1$ the exact profile~\eqref{eq:dTfixed} indeed solves~\eqref{eq:linressc}, yielding $\nu=\frac{1}{2}$.

For free BCs with $\lambda \neq \omega$ we have $0 < R < 1$ and~\eqref{eq:linressc} has to be solved numerically, for lack of an exact solution. The resulting continuum of kernels interpolates between the $R=0$ and $R=1$ curves.
A striking feature is that the singularity exponent $\nu$ is observed to depend continuously on $R$~\cite{lepri_p2011}.

\begin{figure}[t]
	\begin{center}
		\includegraphics[width=0.85\textwidth]{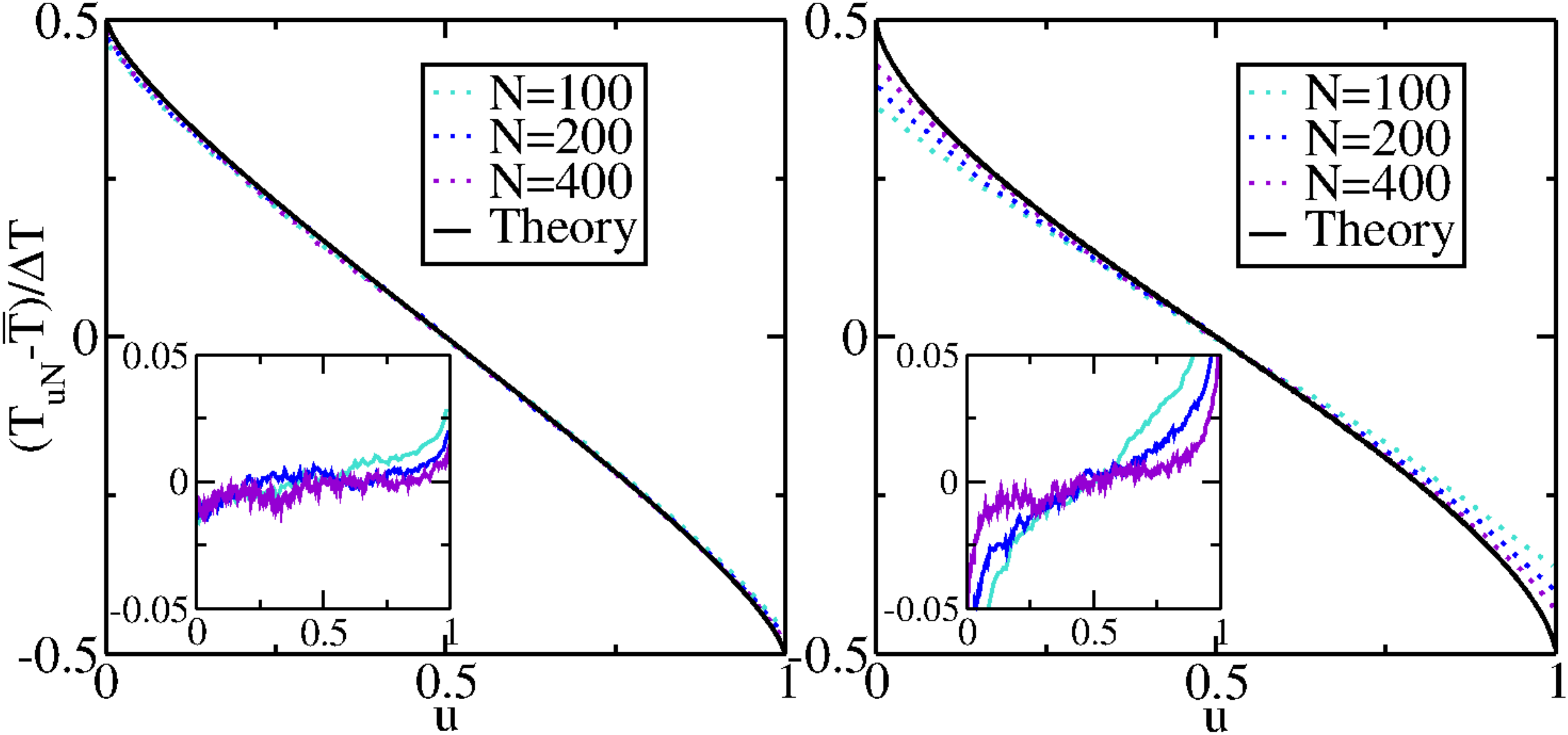} 
	\end{center}
	\caption{\small Rescaled temperature profile for resonant BCs (left) and free BCs with $R=\frac{1}{2}$ (right). In the main plots results of Monte-Carlo simulations for increasing system sizes $N=100$, $200$ and $400$ are compared to the theoretical predictions given by~\eqref{eq:Tpsonin} with $\alpha=\frac{1}{2}$ and the numerical solution of~\eqref{eq:linressc}-\eqref{eq:solK} with $R=\frac{1}{2}$, respectively. In the insets the differences between measurements and theory are shown. The other parameter values are $\Tp=1.5$, $\Tm=0.5$, and $\omega=\gamma=1$.}
	\label{fig:tprof}
\end{figure}

To determine the exponent $\nu$ we derive~\eqref{eq:linressc} with respect to $u$. This derivative should identically vanish. Assuming that $\cTp(u) \sim u^{-\nu}$ for small $u$, the derivative of~\eqref{eq:linressc} exhibits a singular term $\propto u^{-\nu-1/2}$, whose coefficient is determined by the singularities of the kernel on the line $u=v$ and at the points $(u,v)=(0,0)$ and $(1,1)$. Eventually, requiring the coefficient of the singular term to vanish provides a condition on $\nu$,
\begin{equation}
\label{eq:nuimpl}
\int_{0}^1 \frac{w^{-\nu}-w^{\nu-1/2}}{(1-w)^{3/2}} \dd w = R \int_{0}^1 \frac{w^{-\nu}+w^{\nu-1/2}}{(1+w)^{3/2}} \dd w.
\end{equation}
Solving~\eqref{eq:nuimpl} using Mathematica we obtain $\nu$ as a function of $R$, which we verify against direct numerical solution of~\eqref{eq:linressc}-\eqref{eq:solK} in Fig.\,\ref{fig:Rnu}. Note that this expression differs form the linear function conjectured in~\cite{lepri_p2011} for $\nu(R)$ based on numerical evidence, although it is quite close to it.

In fig.~\ref{fig:tprof} we plot the temperature profiles for two cases: free resonant BCs and free BCs with $R=\frac{1}{2}$. 
In all cases the profiles predicted by~\eqref{eq:linressc}-\eqref{eq:solK} are in good agreement with Monte-Carlo simulations of the HCME and therefore validate the above analysis.

This work arose from the incompatibility between the temperature profile of the HCME predicted by the linear response relation~\eqref{eq:linres} using the infinite system kernel and the exact solution for fixed BCs~\eqref{eq:dTfixed}. In all considered cases it turns out that the linear response relation~\eqref{eq:linres} can be applied provided that the proper BCs are used to derive the kernel. For the HCME the HD theory can be supplemented with BCs~\eqref{eq:bcsfree} for the HD equations, derived from the microscopic BCs. The linearity of the HD equations allows for an exact analytical treatment in the large $N$ limit, which is fully consistent with the previously known solution for the fixed BCs~\eqref{eq:dTfixed} and numerical evidence.

An important result of the present study is the clear analytical and conceptual connection between the L{\'e}vy flight picture~\cite{lepri_p2011, dhar_s_d2013} and the HCME. 
Equation~\eqref{eq:linressc} with the kernel~\eqref{eq:solK} is the same as that obtained for the density profile of a system of particles performing L{\'e}vy flights with a length distribution decaying  
with a power of $\frac{5}{2}$~\cite{lepri_p2011,dhar_s_d2013}.
In this model the L{\'e}vy particles are reflected with probability $R$ at the boundaries. 
In the HCME the fluctuations of the sound modes are identified with the quanta of heat being carried by the L{\'e}vy flyers and the spreading of the peaks corresponds to the decrease of the flight length probability with distance.

The linear response relation~\eqref{eq:linres} should be valid for other systems when the temperature difference is small, $\Delta T \ll \Tb$ and the system size $N$ is large.
In this regard, the HCME constitutes a landmark in which HD BCs can be obtained, the HD equations~\eqref{eq:hydro}-\eqref{eq:bcsfree} can be solved exactly and the solution is valid for any $\Delta T$. The analysis leading to~\eqref{eq:nuimpl} shows that a single reflection of the sound peaks can be enough to modify the asymptotic temperature profile.
It would be interesting to apply the same approach to other systems and obtain the temperature profiles for systems such as Fermi-Pasta-Ulam chains or gas models. We defer this study to later publication~\cite{longpaper}.

\section*{Acknowledgments}

	AK would like to acknowledge the hospitality of the Weizmann Institute of Science where part of the work was done while he was visiting. The support of the Israel Science Foundation (ISF) is gratefully acknowledged.

\section*{References}

\appendix
\section{Derivation of the kernels}
\label{section:appkernel}

We start from the solutions of the HD equations, Eq.\,\eqref{eq:phipm}.
We first average~\eqref{eq:phipm} over the realizations of the noises $\etap$ and $\etam$. We get, for $\sigma=\pm$,
\begin{eqnarray}
\langle \phi_\sigma(x,t)^2 \rangle_{\etap,\etam} &=& \left[ \int_{\xi=0}^N \dd \xi (f_{\sigma,+}(x,\xi,t) \phip(\xi) + f_{\sigma,-}(x,\xi,t) \phim(\xi)) \right]^2 \\
&& + 2 D \int_{\xi=0}^N \dd \xi \int_{t'=0}^t \dd t' [(\partial_{\xi} f_{\sigma,+}(x,\xi,t-t'))^2 \nonumber \\&&\hspace{4cm}+ (\partial_{\xi} f_{\sigma,-}(x,\xi,t-t'))^2 ],\nonumber
\end{eqnarray}
where we performed integration by parts in the second term and we defined $\phi_\sigma(\xi) = \phi_\sigma(\xi,0)$ for short. The terms on the second line cancel when computing the current $J = G (\phip^2-\phim^2)$~(\eqref{eq:defJ}). The other terms give
\begin{eqnarray}
\langle J(x,t) \rangle_{\etap,\etam} &=& G \int_{\xi=0}^N \dd \xi \int_{\zeta=0}^N \dd \zeta \bigg[  (f_{+,+}(x,\xi,t) \phip(\xi) + f_{+,-}(x,\xi,t) \phim(\xi)) \nonumber \\ &&\times (f_{+,+}(x,\zeta,t) \phip(\zeta) + f_{+,-}(x,\zeta,t) \phim(\zeta)) \nonumber \\ 
&& - (f_{-,+}(x,\xi,t) \phip(\xi) + f_{-,-}(x,\xi,t) \phim(\xi)) \nonumber \\ &&\times (f_{-,+}(x,\zeta,t) \phip(\zeta) + f_{-,-}(x,\zeta,t) \phim(\zeta))\bigg].
\end{eqnarray}

We now have to average over the initial condition $\phipm(\xi)$. At initial time the system is in thermal equilibrium, so we assume the fields to be Gaussian and uncorrelated in space,
\begin{equation}
\label{eq:eqcorr}
\langle \phi_\sigma(\xi) \phi_\tau(\zeta) \rangle_\mathrm{eq} = S \delta_{\sigma,\tau} \delta(\xi-\zeta),
\end{equation}
and higher-order correlations can be expressed as combinations of two-point correlations~\eqref{eq:eqcorr} using Wick's theorem. In order to obtain $\langle J(x,t) J(y,0)\rangle_\mathrm{eq}$ we need to compute
\begin{eqnarray}
\label{eq:fourpt}
&&\langle \phi_\sigma(\xi) \phi_\tau(\zeta) (\phip^2(y)-\phim^2(y))\rangle_\mathrm{eq} = 2 S^2 \delta_{\sigma,\tau} [\delta_{\sigma,+}-\delta_{\sigma,-}] \delta(\xi-y) \delta(\zeta-y),
\end{eqnarray}
where two potentially infinite terms proportional to $\delta(0)$ have canceled. For the current-current correlation we obtain
\begin{eqnarray}
\langle J(x,t) J(y,0)\rangle_\mathrm{eq} &=& 2 G^2 S^2 \big[ f_{+,+}(x,y,t)^2 + f_{-,-}(x,y,t)^2 \nonumber \\ &&-f_{+,-}(x,y,t)^2 -f_{-,+}(x,y,t)^2\big].
\end{eqnarray}
The next step is integration over time. To simplify the calculations, we consider the large $N$ regime with $u = \frac{x}{N}$ and $v = \frac{y}{N}$ fixed. Using the explicit expression of the Green functions, \eqref{eq:greenf}, we have
\begin{eqnarray}
\label{eq:JJ}
&&\int_{t=0}^\infty \dd t \langle J(Nu,t) J(Nv,0) \rangle \nonumber \\
&=& 2 G^2 S^2 \int_{t=0}^\infty \dd t \sum_{\sigma,\tau = \pm} \sum_{m,n=-\infty}^{\infty} \sigma \tau \frac{r^{2 (n+m) + \sigma-\tau}}{4 \pi D t}  \\
&&\times \exp\left(-\frac{(N(u - \sigma \tau v  + 2 \sigma n)   -\sigma \cs t)^2+(N (u- \sigma \tau v + 2 \sigma m ) -\sigma \cs t)^2}{4 D t}\right) \nonumber \\
&\simeq&  \frac{G^2 S^2}{D \pi} \sum_{\sigma,\tau = \pm} \sigma \tau \sum_{n=-\infty}^{\infty}   r^{4 n + \sigma-\tau} \ee^{\frac{\sigma \cs N}{D}  (u +2 n \sigma- v \sigma \tau)} K_0 \left( \frac{\cs N |u+ 2 n \sigma - v \sigma \tau|}{D} \right) \nonumber \\
&\simeq&  \frac{G^2 S^2}{D \pi} \sqrt{\frac{\pi D}{2 N \cs}}\bigg[ \left( \frac{\Theta(u-v)}{\sqrt{u-v}} + \sum_{n=1}^\infty \frac{r^{4 n}}{\sqrt{2 n +u - v}} \right) - \left( \sum_{n=0}^{\infty}   \frac{r^{4 n + 2}}{\sqrt{2 n + u + v}} \right) \nonumber \\
&& - \left( \sum_{n=1}^{\infty}   \frac{r^{4 n - 2}}{\sqrt{2 n - u - v}}\right) + \left(\frac{\Theta(v-u)}{\sqrt{v-u}} + \sum_{n=1}^\infty \frac{r^{4 n}}{\sqrt{2 n -u + v}}\right)
\bigg]. \nonumber 
\end{eqnarray}
In the second line we have neglected terms with $m \neq n$, that are exponentially small in $N$, and the integral over $t$ has been computed in terms of the Bessel function $K_0$. In the large $N$ limit the argument of the Bessel function becomes large. From the asymptotic form $K_0(z) \sim \sqrt{\frac{\pi}{2 z}}\ee^{-z}$ we get that terms with $\sigma u +2 n - v \tau < 0$ are exponentially small in $N$. As a consequence, the sum over $n$ can be restricted to $n \geq 1$ for $(\sigma,\tau) = (-,+)$ and to $n \geq 0$ for the three other values of $(\sigma,\tau)$. After replacing the Bessel functions with their asymptotic forms we obtain the third line of~\eqref{eq:JJ}, where each parenthesis corresponds to a value of $(\sigma,\tau)$. Note that the two terms for $n=0$ and $(\sigma,\tau) = (+,+)$ and $(-,-)$ contribute only for $u>v$ and $u<v$, respectively, and combine to give a term proportional to $|u-v|^{-1/2}$. From the third line of~\eqref{eq:JJ} one simply has to shift the index $n$ in the second sum to obtain the result~\eqref{eq:solK}.

\end{document}